\documentclass{PoS}

\title{Light Baryons from 2+1 flavor DWF QCD}

\ShortTitle{Light Baryons}

\author{\speaker{C.M. Maynard}\\
        EPCC, School of Physics and Astronomy, University of
        Edinburgh, EH9 3JZ\\
        E-mail: \email{c.maynard@ed.ac.uk}}

\author{For the RBC and UKQCD collaborations}

\abstract{We present results from the RBC and UKQCD 
collaboration ensembles of 2+1 flavor DWF QCD
for the light baryon spectrum.
}

\FullConference{The XXVII International Symposium on Lattice Field Theory - LAT2009\\
		 July 26-31 2009\\
		 Peking University, Beijing, China}

\begin{document}

There are many reasons to study the light Baryons some of which are
summarised by the late Nathan Isgur in ``Why $N^{\star}$'s are important''~\cite{Isgur:2000ad}, which I
paraphrase here:
\begin{itemize}
   \item {\em Nucleons are the stuff of which world is made.}
   \item {\em They are the simplest system in which the
       quintessentially nonabelian character of QCD is manifest.
       There are $N_c$ quarks in proton because there are $N_c$
       colours.}
   \item {\em Baryons are sufficiently complex to reveal physics hidden from us in Mesons.}

\end{itemize}
Indeed, even the low lying spectrum is not completely understood. For
example the nature of the Roper N(1440) $J^P = \frac{1}{2}^+$ resonance is a long standing
puzzle. This state doesn't naturally fit into the quark models. In most
models, the parity of the excitations of the nucleon are
alternatively negative, then positive. The Roper has positive parity,
and lies between Nucleon ground state N(939) $J^P = \frac{1}{2}^+$ and the negative parity
excitation  N(1535) $J^P = \frac{1}{2}^-$. In the quark models, the
positive parity excitation lies above the negative parity
excitation and this has lead to speculation that the Roper is not an
excitation of the nucleon but some other state. In principle lattice
QCD can definitively resolve this issue simply by determining the
spectrum of excited nucleons. However, excited states are difficult,
requiring both large volumes and high statistics.

The preliminary results presented in these proceedings are from the RBC and UKQCD 2+1 flavor Domain Wall Fermion (DWF) data sets, with the Iwasaki Gauge action and the size of the fifth dimemsion, $L_S=16$. The details are shown in table~\ref{tab:datasets}, the $24^3$ data is described in~\cite{Allton:2008pn} and the $32^3$ data in~\cite{bob,CK}.  DWF have lattice versions of the continuum QCD symmetries, in particular flavor, chiral and the lorentz (e.g. parity) symmetries. These symmetries come at an increased cost of simulation, but they protect matrix elements from mixing with other operators.  This makes renormalisation of matrix elements simple\footnote{or at least, simpler.}. So the target quantities for DWF calculations are matrix elements. Of course, one can use the same ensembles for spectrum calculations. Symmetry is also good for the spectrum, but what is really required is very high statistics, and what has been paid for with DWF is symmetry. In this sense, spectrum calculations are the poor relation in DWF calculations. Moreover, the smearings and sources tuned for the matrix element calculations are not
optimal for the spectrum, and so each ensemble has different sources and numbers of sources.

 \begin{table} \begin{center} \caption{Ensemble details}\label{tab:datasets} \begin{tabular}{cccccc} $\beta$ & V & $a^{-1}$ (GeV$^{-1}$) & $am_f(s)$ & $am_f(ud)$ & $am_{\rm res}$ \\\hline $2.13$ & $24^3$ & $1.75(3)$ & $0.04$ & ${0.03,0.02,0.01,0.005} $ & $
      0.0030(1) $ \\
      $2.25 $ & $32^3$ & $2.33(4)$ & $0.03$ & $ {0.008,0.006,0.004} $ & $0.00067(1)$ \\\hline \end{tabular} \end{center}\vspace{-0.5cm} \end{table}

\begin{figure}\vspace{-0.5cm}
\begin{center}
\includegraphics[width=.65\textwidth]{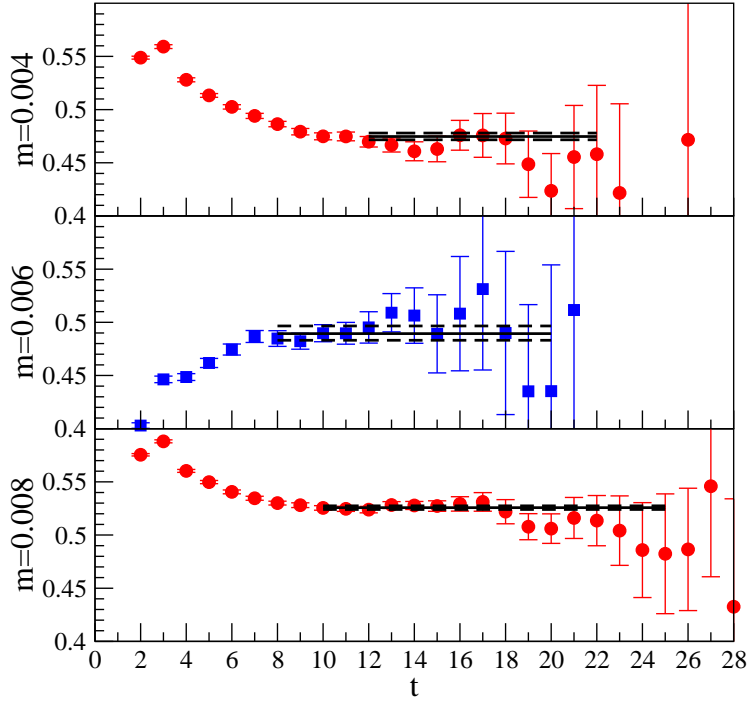} 
\caption{Effection mass of the nucleon correlator for the $32^3$ ensembles. The y-axis labels denote $am_f(ud)$. The upper and lower panels (circles) show correlators generated by the LHPC collaboration with Gaussian sources, the middle panel (squares) shows correlators generated with wall sources. The horizontal lines show the fit of an exponential to the correlator, and the span denotes the fit ranges.}
\label{fig:effMass}
\end{center}\vspace{-1cm}
\end{figure}

Shown in Figure~\ref{fig:effMass} are the effective masses for the $32^3$ data. As different sources were used to generate the data, the quality of the plateaux depends of the source used. In particular the $m_f(ud)=0.006$ was generated with a wall source and the data has a plateau which approaches from below. It is not clear the extent of the ground state saturation before noise swamps the signal. Certainly, the quality of the wall source data is inferior to the Gaussian source data. The results for the nucleon mass on the $24^3$ data were presented at the previous lattice conference~\cite{Blum2008aa}.

\begin{figure}\vspace{-0.5cm}
\begin{center}
\includegraphics[width=0.6\textwidth]{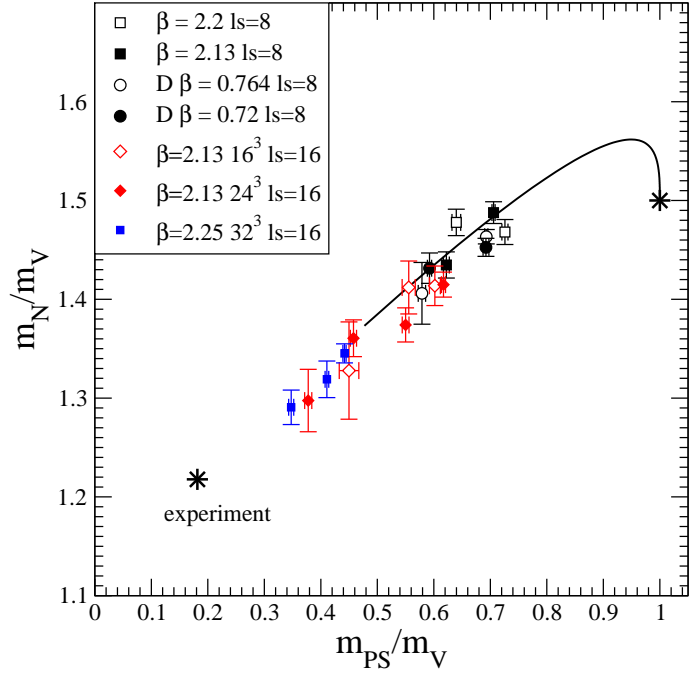}
\caption{The Edinburgh plot for all the DWF data generated on QCDOCs and at the Argonne Leadership Computing Facility, IBM BG/P ($32^3$ data). The solid line shows a quark model prediction}\label{fig:edplot}
\end{center}\vspace{-1cm}
\end{figure}

Ultimately the chiral extrapolation, to determine the spectrum at the physical quark masses, will need to be attempted, but its instructive to compare the nucleon masses of different simulations. This is done
using the Edinburgh plot, and it is a good consistency test of a calculation. Shown in Figure~\ref{fig:edplot} is the Edinburgh plot for all the 2+1 flavor DWF data. There are two things to note about this plot. Data from Lattice QCD now goes beyond the end of the quark model line. The data comes from different simulations and with changes to gauge action (DBW2 and Iwasaki), gauge coupling and hence lattice spacing, lattice volume and quark mass.
At the level of statistical resolution the data lie on (apparently) a universal curve. One may conclude from this that the DWF calculations are internally consistent and these data are plotting the QCD curve.

\begin{figure}
\begin{center}
\includegraphics[width=0.85\textwidth]{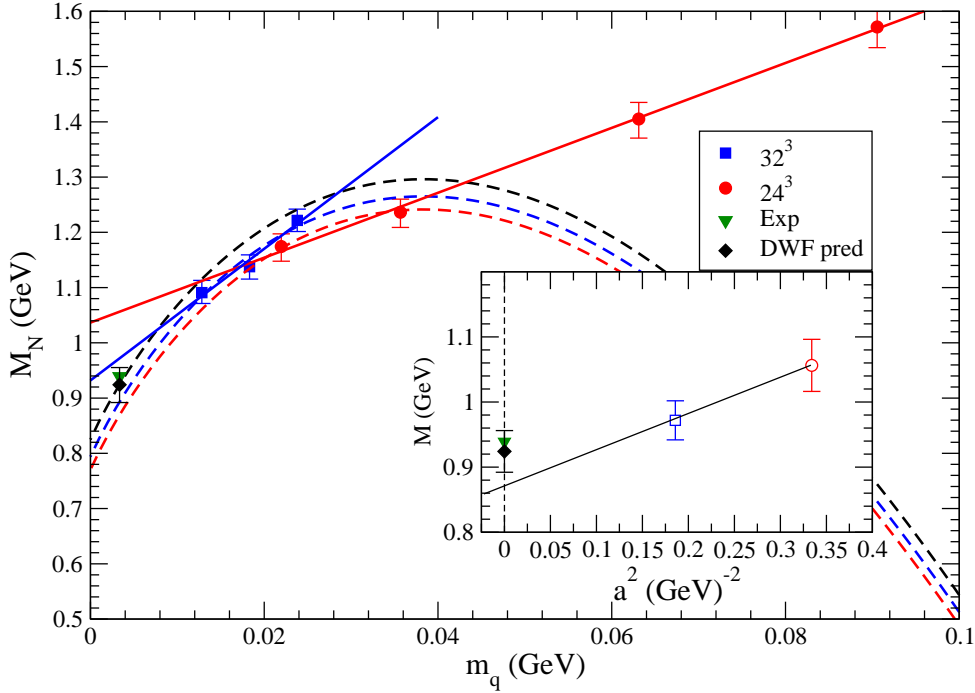}
\caption{The nucleon mass versus renormalised quark mass in physical units. Blue (red) symbols denote $32^3$ ($24^3$) data. The straight lines show linear extrapolations at fixed lattice spacing. The curves show the chiral fit (equation~(2)). 
The black symbol is the DWF prediction, and the green
triangle is the experimental result. The inset shows the scaling behaviour of the linear extrapolation, and compares to experiment (green traingle) and the chiral fit (black diamond). \label{fig:NucChiral}
}
\end{center}\vspace{-1cm}
\end{figure}

Many calculations now include sophisticated chiral extrapolations involving partially quenched data, with many different valence quark masses, including one on these data sets~\cite{bob,CK}. Only unitary data, that is, the mass of the quarks in the sea is the same as the valence, is available for the baryon spectrum. Shown in Figure~\ref{fig:NucChiral} are the nucleon masses. The simplest method for obtaining the continuum and chiral result is a linear extrapolation to the physical quark masses, and then ``extrapolate''\footnote{Naively drawing a straight line through two points.} the data to the continuum limit. However, more sophisticated chiral behavior can be examined. This is motivated by heavy baryon chiral perturbation theory~\cite{Jenkins:1991es}, where the nucleon mass is given as a function of the pion mass squared.

\begin{equation}
\label{eq:HBCPT}
  M_N = M_0 - 2\alpha m_{\pi}^2 - \frac{3 g_A^2}{4 \pi f} m_{\pi}^3 + {\rm logs}
\end{equation}

This equation is used to suggest a form for a chiral fit for the
nucleon mass at both lattice spacings, including quadratic lattice artifacts.

\begin{equation}
M_N = c_0 + c_1 m_q + c_2 m_q^{3/2} + c_3a^2
\label{eq:chiralFit}
\end{equation}

This equation is then fitted to the five lightest data, $am_f = \{0.004,0.006,0.008\}$ from the $32^3$ ensembles and $am_f = \{0.005,0.01\}$ from the $24^3$ ensembles simultaneously. This fit is also shown in 
Figure~\ref{fig:NucChiral}. This a four parameter fit to five data and has an uncorrelated $\chi^2/{\rm dof} \sim 1.3 $, which is perhaps higher than one would like for an uncorrelated fit. The nucleon mass then
predicted is 
\begin{equation}
  M_N = 0.924(30) {\rm GeV}
\end{equation}
This error is statistical only as these are preliminary results, but does agree with the experimentally measured value.

How sensible is this fit? Firstly, the curves plotted in Figure~\ref{fig:NucChiral} are drawn using the
fit parameters with the lattice spacing set to its $24^3$ value in red, $32^3$ value in blue, and zero in
black. The gradient of these curves is similar to the gradient of the straight line fits as they pass through
the data. Whilst the gradient of the lattice spacing term has the opposite sign to the gradient of the continuum extrapololation shown in the inset, they are both relatively small, $c_3 \sim -0.17(5)$ for the 
chiral fit, $\sim 0.57$ for the linear fits. So lattice artefacts are small, again this can be seen in the
plot as the data from different lattice spacings is close to each other. So, despite the large $\chi^2/{\rm dof}$, the fit reflects the data reasonable well. Moreover, the coefficent of the $m_q^{3/2}$ term, $c_2$ can
be related to $g_A$ once a fudge factor to convert from renormalised quark mass to pion mass squared is included. The value of $g_A$ obtained from the fit is then $g_A^{\rm fit} \sim 2.2$. This can be compared to the experimental value $g_A^{\rm exp} \sim 1.27$. This is clearly wrong, but critically, it is the right order of magnitude and {\em the right sign}. So, the fit seems reasonable given the data.

Besides high statistics, large volumes are required for the baryon spectrum as these states are physically
big, certainly compared to mesons. Finite size effects (FSE) are therefore an important source of systematic
error. Several baryonic quantities have been measured and reported on these data sets. In~\cite{Yamazaki:2008py}, a large FSE is reported for the $24^3$ data for the axial charge of the nucleon.
However, in~\cite{Aoki:2008ku} no FSE was observed in the nucleon spectrum between the the $24^3$ data 
and a smaller volume of $16^3$ with all other quantities held fixed, for $am_f=0.01$. The statistical precision of the data is around $1-2\%$. Subsequently, the lightest $24^3$ ensembles have been extended, and it is these extended ensembles have been included in this work. The FSE is analyised for the heaviest
quark mass in the chiral fit. However, it is instructive to compare the sizes of the lattice in terms of
the Compton wavelength of the pion, $m_\pi L$. For the $16^3$ and $24^3$ data used to estimate FSE, the 
$m_\pi L$ values are $\sim 3.9$ and $\sim 5.6$ respectively. For the lightest datum analysed in this work,
the $32^3$ $am_f=0.004$, the $m_\pi L$ value is $\sim 4.1$. This suggests that for the nucleon mass at least,
FSE should be less than, say $1\%$. This is in accordance with ``Lattice folklore'' from other studies, that the FSE for the nucleon mass should be less than $1\%$ for an $m_\pi L$ value of $> 4$.  

\begin{figure}
\begin{center}\vspace{-0.5cm}
\includegraphics[width=0.6\textwidth]{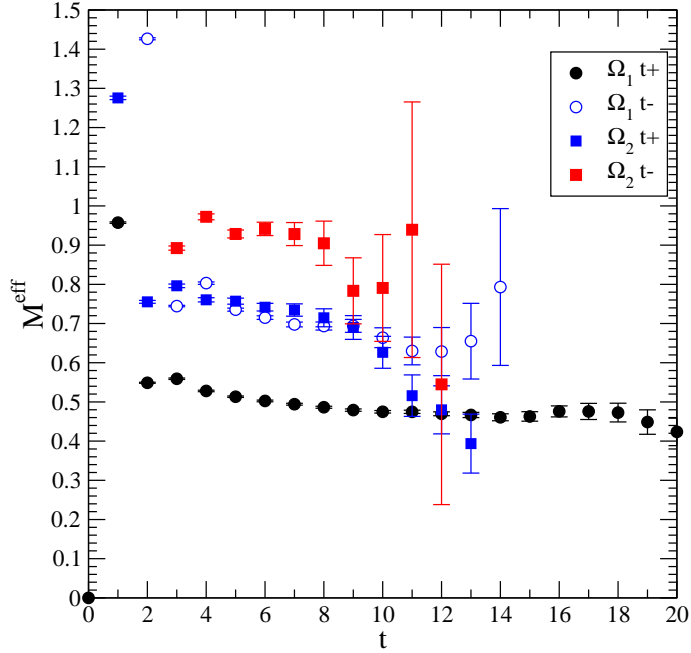}
\caption{The effective masses of correlation functions constructed from the operators
defined in equations~(4-5) with quark mass $am_f=0.004$. The label $t_+$ denotes the
forward propagating state, and $t_-$ denotes the time reversed backward moving state.\label{fig:NdoubleStar}
}
\end{center}\vspace{-1cm}
\end{figure}

The nucleon operators used to construct the correlation functions are
\begin{eqnarray}
\label{eqn:NucOps}
\Omega_1 & = & \left(\psi C \gamma_5 \psi\right)\psi \\
\Omega_2 & = & \left(\psi C  \psi\right)\psi 
\end{eqnarray}
On a lattice with anti-periodic boundary conditions, the forward propagating state of the correlation function constructed from the $\Omega_1$ operator projects onto the positive parity nucleon, and the backward propagtor which has the opposite parity projects onto the negative parity excitation, the $N^\star$ state.
The $\Omega_2$ operator has negative parity, and so the forward propagator of its correlation function
projects onto a negative parity state and can be used as an additional estimate of the $N^\star$. 
The effective masses of these correlation functions are shown in Figure~\ref{fig:NdoubleStar}. The blue symbols show the effective masses for the negatice parity state, showing good agreement.
Also shown (in red) is the backward propagator of $\Omega_2$ correlation function. What state does this project onto? It has positive parity, and is clearly not the nucleon. It is possible that this is the positive parity excitation of the nucleon. This state clearly lies above the negative parity excitation.

\begin{figure}
\begin{center}
\includegraphics[width=0.8\textwidth]{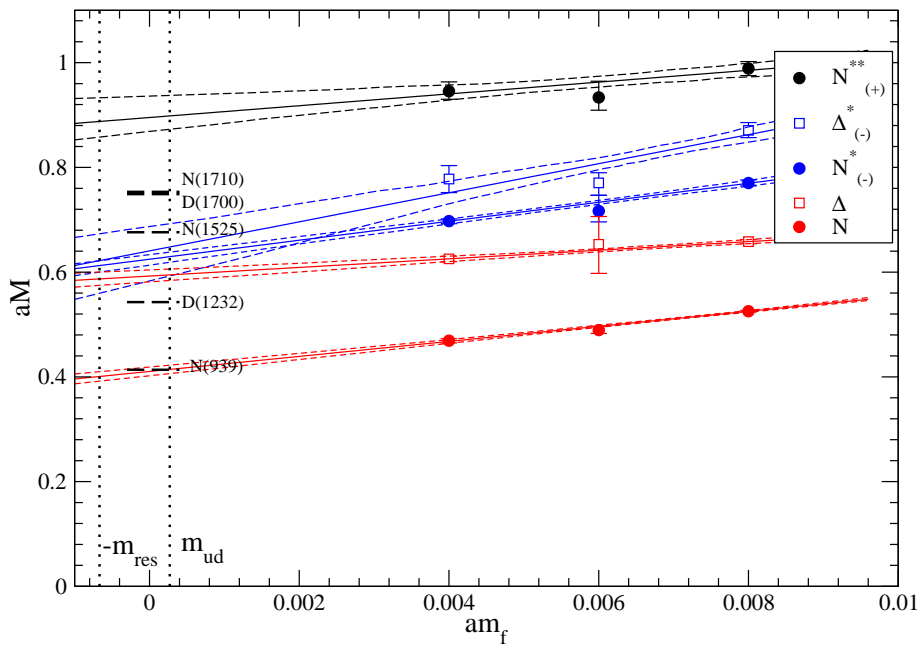}
\caption{Baryon masses in lattice units versus bare quark mass $am_f$. $32^3$ data only. Dotted vertical lines show the light quark mass and the chiral limit, labelled $m_{ud}$ and $-am_{\rm res}$ respectively. Dashed horizontal lines show the physical spectrum, in lattice units.\label{fig:Baryons}
}
\end{center}\vspace{-1cm}
\end{figure}

Shown in Figure~\ref{fig:Baryons} is the light baryon spectrum for the $32^3$ data. There are several limitations to this result, only a naive linear extraploation, only one lattice spacing, no estimate of FSE for the higher states, which could be potentially severe, and poor plateaux for the $am_f=0.006$ wall source data. However, it is interesting to compare to the physical spectrum, shown in lattice spacing units. The nucleon, delta and $N^{\star}$ agree reasonably well with experiment. The higher
excitations, the $\Delta^{\star}$ and the $N^{\star\star}$ are more speculative, as they are more likely to be
effected by the quality of plateaux and FSE. Considering the ordering of states
only, if the backward moving $\Omega_2$ operator is indeed the positive parity excitation of the nucleon
then this would exclude the N(1440) Roper resonance from being a nucleon.

We have studied the light baryon spectrum for the 2+1 flavor DWF QCD ensembles and explored some of the
issues necessary to achieve the result. We present a preliminary result for the nucleon mass using
a combined chiral/continuum fit which agrees with experiment. We also present a more speculative
result for the excited spectrum, which taken at face value suggests that the Roper resonance is not
a nucleon, but there several systematic uncertainties which are not sufficiently controlled to make this a concrete result.

We thank all members of the UKQCD and RBC collaboration. Computations were performed 
on the QCDOC machines at the University of Edinburgh and Columbia University, the US DOE 
and RBRC facilities at the Brookhaven National Laboratory and at the Argonne Leadership Class 
Facility. 


\end{document}